\documentstyle[aps,epsfig,multicol]{revtex}

\begin{document}

\bibliographystyle{prsty}

\title{Wigner-Dyson Statistics from the Keldysh $\sigma$-Model}

\draft

\author{Alexander Altland$^{1}$, and Alex Kamenev$^{2}$}

\address{$^{1}$ Theoretische Physik III, Ruhr-Universit\"at
  Bochum, 44780, Germany.\\
  $^{2}$ Department of Physics, Technion, Haifa 32000, Israel.
  \\
  {}~{\rm (\today)}~
  \medskip \\
  \parbox{14cm} {\rm The level statistics of disordered metallic
    grains with broken time reversal invariance is obtained from a
    saddle point analysis of the Keldysh nonlinear $\sigma$--model.
    \smallskip\\
    PACS numbers: 71.23.-k, 71.45.-d, 73.23.-b}\bigskip \\ }

\maketitle

\begin{multicols}{2}
\narrowtext 

In the past two decades, the field theory of disordered electronic
systems, a matrix version of the nonlinear $\sigma$--model, has
attracted a great deal of theoretical attention
\cite{Efetov,Wegner79,Kamenev98,Andreev95,Kamenev99,Kamenev00}.  The
$\sigma$-model provides the perhaps most rigorous and comprehensive
theoretical framework by which interacting and non--interacting
disordered electronic systems can be explored.  Three different,
albeit closely related versions of this model exist: The
supersymmetric (SUSY) \cite{Efetov}, replica \cite{Wegner79} and
dynamic or Keldysh \cite{Kamenev98} $\sigma$-model.  Due to different
microscopic starting points these three formulations cover partly
complementary application areas: whereas SUSY is tailor made to the
analysis of non-perturbative phenonmena, the replica, and in
particular the Keldysh formalism are strong at problems involving
interactions -- about which SUSY has nothing to say. This
constellation amounts to some theoretical vacuum because sooner or later
non-perturbative interaction related problems will come into focus.

An important step towards improving the situation was made
recently\cite{Andreev95} (first within the framework of SUSY) when it
became clear that the large energy asymptotic of Wigner-Dyson (WD)
level statistics -- {\it the} non-perturbative prototype problem --
may be obtained from a careful analysis of the saddle point structure
of the $\sigma$--model (in contrast to a full integration over the
field manifold which is out of question for the non-SUSY
representants).  Subsequently these ideas were adopted to the replica
$\sigma$--model \cite{Kamenev99}, which proved to be useful for the
solution of simple interacting problems \cite{Kamenev00}. Yet, the
application spectrum of the replica theory, excluding non-equilibrium
phenomena, is significantly narrower than that of its Keldysh
counterpart. Furthermore, replica analyses categorically rely on a
formidable analytic continuation procedure, which tends to obscure the
underlying phyiscs and essentially complicates practical applications.
It is therefore desirable to formulate a prescription whereby
non-perturbative information can be obtained from the conceptually
more transparent Keldysh formalism. This is the subject of the present
Letter.

All the theories mentioned above share the normalization condition
$Z=1$, where $Z$ is the functional partition function. It is this
universal normalzation property which makes the models useful in the
analysis of disorder properties\cite{Efetov}. The normalization in
turn is due to a global internal symmetry underlying the model:
supersymmetry, replica permutation symmetry or, within the Keldysh
formulation, a symmetry principle related to the causality of the
theory. Building on these structures, the saddle point analysis
of any $\sigma$--model must consist of three parts: (i) identification
of {\em all} saddle points and their action, (ii) proof that $Z=1$
holds within the stationary phase scheme (iii) saddle point evaluation
of any specific correlation function.  Using WD spectral statistics as
a example we are going to apply this program to the Keldysh model of
unitary (broken time reversal invariance) symmetry.  The
generalization to different symmetries and observables will be
discussed elsewhere \cite{ak}.

We wish to compute the two-point correlation function
$R(\epsilon_1,\epsilon_2)= \Delta^2 \langle \langle \rho(\epsilon_1)
\rho(\epsilon_2) \rangle \rangle = \Delta^2/(2\pi^2) {\, \Re\,}
\langle\langle {\,\rm tr\,} \{G(\epsilon_1^+)\}{\,\rm
  tr\,}\{G(\epsilon_2^-)\} \rangle\rangle$ of the density of states
$\rho(\epsilon)$ for a model of free electrons with broken time
reversal invariance and subject to a random potential. Here $\Delta$
is the mean level spacing and $\langle \langle \dots
\rangle \rangle$ stands for a (cumulative) disorder average.  Our
starting point is the low energy effective partition
function~\cite{Kamenev98} $Z= \int {\cal D}Qe^{iS[Q]}$, where
\begin{equation}
  \label{eq:1}
  i S[Q] = -{\pi \over 4 \Delta } \int\!\!
            {d^d{\bf r}\over L^d} {\,\rm tr\,}\left\{D {\bf \partial} Q
    {\bf \partial} Q + 4i \hat \epsilon Q\right\},
\end{equation}
$D$ is the diffusion constant and $L$ the system size.
The matrix--valued field $Q=\{Q_{\epsilon,\epsilon'}^{l,l'}\}$ acts
in a product space defined through an index $l=1,2\;$ labelling the
forward and backward Keldysh time contour, ${\cal
  C}^l=\{t \in (\mp\infty,\pm\infty)\}$, and the energy indices,
$\epsilon$, Fourier conjugate to the time variables $t$ on ${\cal
  C}^l$.  For all what follows it will be convenient to discretize the
(a priori continuous) energy variables in units of some small spacing
$\delta_\epsilon$.  Since the application range of the effective
action is limited to energies $|\epsilon| < \tau^{-1}$ 
\cite{Wegner79,Kamenev98}, this
manipulation implies that the $Q$'s become matrices of finite
dimension $2 K$, where $K = 2/(\delta_\epsilon \tau)$.  The internal
structure of these matrices is defined through (a) the constraint
$Q^2({\bf r})=\openone$ and (b) Hermiticity, $Q^\dagger = Q$.
Finally, $\hat \epsilon = {\,\rm diag\, }(\hat \epsilon^1,\hat
\epsilon^2)$ is a $2K$-dimensional diagonal matrix where $\hat
\epsilon^l= {\, \rm diag } (\epsilon_1,\dots \epsilon_n, \ldots
\epsilon_K)-i(-1)^l 0$.

The definition of the low energy effective action (\ref{eq:1}) implies
one more important structural element which is not explicit in the
notation: by definition of the trace 'tr', the energy
integration/summation over any continuous and analytically benign
function of the energy indices $\epsilon$ produces zero (e.g. ${\,\rm
  tr\,}\{\hat \epsilon\}=0$).  For lack of better terminology we will
refer to this feature as 'causality'.  That the causality criterion is
not explicit in the definition of the trace is not just a matter of
notational convenience. The crux is that to rigorously retrieve the
full causality properties of the microscopic Keldysh partition
function, {\it all} energies, $\epsilon$, including energies in excess
of the width of the spectrum of the microscopic Hamiltonian, have to
be taken into account.  The philosophy behind declaring the causality
principle to an intrinsic feature of the effective action (\ref{eq:1})
is that $S$ merely represents the low energy sector of some larger
parent theory, $S+S_{\rm high}$. After the inclusion of the high
energy sector $S_{\rm high}$, all energy summations could, in
principle, be extended to infinity and the correct spectral structures
of the Keldysh partition function would be retrieved.

This anticipation has been at the root of previous effective action
formulations of the Keldysh approach, and, needless to say, is of
relevance for all matters related to spectral statistics. 
We will therefore subdivide our analysis into two parts.
Taking a pragmatic point of view we will first show how the spectral
correlation function can be obtained from the low energy model,
Eq.~(\ref{eq:1}), once the causality assumption has been made.  In a
second part we will then show how causality can be made {\it
  manifest}, on the expense of including large energies. We emphasize
that part II of the analysis has been included for reasons of conceptual
completeness; in practical
applications, the large energy sector will normally not play a role.


{\it Part I:} The simplest way of computing $R(\epsilon_1,\epsilon_2)$
from the functional integral over the effective action $iS[Q]$, is to
couple the energy vector $\hat \epsilon$ to sources. This can be done
by generalizing $\hat \epsilon^l \to \hat \epsilon^{l,\kappa} \equiv
\hat \epsilon^l + \hat \kappa^l$, where the energy diagonal matrices
$\hat \kappa^l= \{ \delta_{nn'} \delta_{n n_l} \kappa_l\}$ and
$\epsilon_{n_l}$ are the discrete energies closest to the arguments
$\epsilon_l$, $l=1,2$.  It is then a straightforward matter to verify
that the definition of the Keldysh partition function implies
$R(\epsilon_1,\epsilon_2) = \Delta^2/(2\pi^2) {\rm
  \,Re\,}\partial^2_{\kappa_1 \kappa_2}\big|_{\kappa=0} Z[\hat
\kappa]-{\rm dis}$, where 'dis' stands for the disconnected part of
the functional average.  To keep the presentation simple we will focus
in the quantitative analysis of this expression on the
contribution of the spatial zero mode $Q({\bf r})\equiv Q={\rm
  const.}$ (which ought to reproduce WD statistics). The inclusion of
the spatially fluctuating modes, which is straightforward and does not
introduce conceptually new elements, will be briefly discussed in the
end.

The key to understanding the structure of the zero mode integration
lies in the observation that the (source-free) effective action $iS[Q]
= - i \pi \Delta^{-1} {\,\rm tr\,}\{\hat \epsilon Q\}$ possesses a
multitude of $2^{2K}$ isolated saddle points. Indeed, any of the
(energy diagonal\cite{foot1}) configurations $\Lambda = {\, \rm
  diag\,}(\pm 1,\pm 1,\dots,\pm 1)$ solves the stationary phase
equation of the model, $\delta_Q S[Q] = \pi \Delta^{-1} [\hat
\epsilon,Q]=0$. In what follows we are going to show that a Gaussian
integration around these saddle points produces WD statistics.

To prepare the integration, consider the contribution of any saddle
point $\Lambda$ with $K+p$ entries $+1$ and $K-p$ entries $-1$. We
first re-order these elements (through some global unitary
transformation) such that $\Lambda$ assumes the form $\Lambda= {\, \rm
  diag\,}(1,\dots,1 ,-1,\dots -1)$. Next, a set of field
configurations weakly fluctuating around $\Lambda$ is introduced
through $Q= T\Lambda T^{-1}$, where the unitary rotation matrices
$T=\exp \Big({{\hphantom -}0^{\hphantom\dagger}\,B\atop -B^\dagger\,
  0}\Big)$, $B$ is a $(K+p)\times (K-p)$-dimensional complex generator
matrix, and the block decomposition corresponds to the signature of
$\Lambda$.

In principle we should now proceed by expanding the action to second
order in $B$ and integrate. Fortunately, however, there is no need to
carry out this program for all $2^{2K}$ saddle points explicitly.  The
reason is that among the entity of saddle points $\Lambda$, there is
one element $\Lambda_0^{l,l'} \equiv (-1)^{l+1}\delta^{l,l'}$ that
plays a distinguished role. What makes $\Lambda_0$ special is that,
unlike the other $\Lambda$'s, its structure is compatible with the
signature of the imaginary increments of $\hat \epsilon$. Building on
this feature, previous analyses of the Keldysh $\sigma$-model indeed
focused on a perturbative expansion around $\Lambda_0$ and did not
take the other saddle points into account.

As a warm-up, let us outline how an integration around the standard
saddle point produces the unit--normalization of the source--free
partition function $Z[\hat \kappa=0]=1$.  Substituting $Q=T\Lambda_0
T^{-1}$ into the zero mode action and expanding to second order in $B$
we obtain the quadratic action \vspace{-0.2cm}
\begin{equation}
\label{eq:2}
i S_{\Lambda_0}^{(2)}[B] = {-i\pi\over \Delta}\Big[{\rm tr} 
\{\hat \epsilon \Lambda_0\} 
-\sum\limits_{n,n'} (\epsilon_{n}^+ - \epsilon_{n'}^-) 
|B_{n n'}|^2\Big].
\end{equation}\\[-0.4cm]
Integration over $B$ then leads to 
$$
\textstyle
Z_0 = {\rm const.}\times e^{-i\pi \Delta^{-1} {\rm tr} 
\{\hat \epsilon \Lambda_0\}} F_0\,   ; \,\, 
F_0 = \prod\limits_{n,n'} (\epsilon_{n}^+  -  \epsilon_{n'}^-)^{-1}  
$$\\[-0.3cm]
as the contribution of $\Lambda_0$ to $Z[0]$.  Due to the causality
property, ${\rm tr}\{\hat \epsilon \Lambda_0\} =0$.  Similarly, $F_0 =
\exp\{-\sum_{nn'} \ln(\epsilon_{n}^+-\epsilon_{n'}^-)\}= \exp
\{0\}=1$, where the presence of the imaginary increments guarantees
that the branch cut singularity of the logarithm is not touched. (We
re-emphasize that, at this stage, the causality rule has the mere
status of a working assumption.  In part II we will make up for its
precise formulation, and show that the proper ultraviolet extension of
the fluctuation determinant fixes the unspecified 'const.' to unity.)
Combining factors, we find $Z_0=1$. As a corollary we remark that
$Z_0=1$ implies vanishing of the total contribution of all {\em
  non}--standard saddle points to $Z[0]$.  Before turning to these
other saddle points, let us discuss the contribution $R_0$ of the
standard saddle point to the spectral correlation function.  A
straightforward expansion of $Z[\hat \kappa]$ to first order in
$\kappa_{1}$ and $\kappa_2$ yields $ R_0 ={1\over 2}\langle \sum_{nn'}
B_{n_1n}B^\dagger_{nn_1} B^\dagger_{n_2n'}B_{n'n_2} \rangle_B $, where
$\langle \dots\rangle_B$ stands for the Gaussian average over the
action (\ref{eq:2}).  Integration over $B$ then leads to $R_0 = 1/(2
s^2)$, where $s=\pi \omega^+ /\Delta$ and $\omega^+ \equiv
\epsilon_{n_1}^+ - \epsilon_{n_2}^-$\cite{foot1a}.

We next turn to the discussion of the other saddle points.  In fact we
will focus on just a single non-standard saddle point $\tilde\Lambda$,
namely the one where the signs of the two entries  corresponding to
our reference energies $\epsilon_{n_{1,2}}$, are flipped:
$\tilde\Lambda \equiv \Lambda_0 - 2\delta^{l1} \delta_{nn_1}
+2\delta^{l2} \delta_{nn_2}$. From the analysis of $\tilde\Lambda$,
the role of all the other saddle points will become clear. The
re-ordering needed to bring $\tilde\Lambda$ into the canonical form
implies that the action $iS_{\tilde\Lambda}^{(2)}$ differs from
$iS_{\Lambda_0}^{(2)}$ in two respects: first, the contribution of the
saddle point itself $iS[\tilde\Lambda]= -i\pi \Delta^{-1}{\,\rm
  tr\,}\{\hat \epsilon \tilde\Lambda \} = i \pi \Delta^{-1} 2\omega^+$
no longer vanishes. Second, in the fluctuation contribution, the two
energy arguments $\epsilon_{n_2}$ and $\epsilon_{n_1}$ are exchanged.
Given these structural changes we find it more convenient to reverse
the order of the evaluation of the functional integral: first
integrate out fluctuations, then expand in the sources. The 
integration over $B_{nn'}$ leads to a fluctuation factor $\tilde F$
similar to $F_0$ above, only that $\epsilon_{n_1}$ and
$\epsilon_{n_2}$ are exchanged and coupled to the respective sources
$\kappa_{1,2}$:\\[-0.5cm]
\begin{equation}
\label{eq:4} 
\tilde F= F_0\, {\epsilon_{n_1} - \epsilon_{n_2}\over \epsilon_{n_2}^\kappa - \epsilon_{n_1}^\kappa}
\prod_{n \not= n_1}
{\epsilon_{n}-\epsilon_{n_2}\over \epsilon_{n} - \epsilon_{n_1}^\kappa}
\prod_{n' \not= n_2}
{\epsilon_{n_1} - \epsilon_{n'} \over \epsilon_{n_2}^\kappa - \epsilon_{n'}}\, .
\end{equation}\\[-0.4cm]
We next add to the products the ``missing'' factors $n = n_{1,2}$ and
use that the now unconstrained products over $n,n'$, as well as the
factor $F_0$, give unity (causality). This leads
to the result\\[-0.6cm]
\begin{equation}
\label{eq:5}
\tilde F= 
{\epsilon_{n_1} - \epsilon_{n_2}         \over  \epsilon_{n_2}^\kappa - \epsilon_{n_1}^\kappa}\;
{\epsilon_{n_1} - \epsilon_{n_1}^\kappa  \over  \epsilon_{n_1}-\epsilon_{n_2} }\;
{\epsilon_{n_2}^\kappa - \epsilon_{n_2}  \over  \epsilon_{n_1} - \epsilon_{n_2} } ={\kappa_1 \kappa_2 \over \omega^{+2} }\, ,
\end{equation}\\[-0.3cm]
where the last equality is valid to leading non-vanishing order in the
sources, $\kappa_1, \kappa_2$.  Notice the proportionality of $\tilde
F$ to $\kappa_1 \kappa_2$. This implies (a) that the sources in the
action $S[\tilde\Lambda]$ may be set to zero (we are differentiating
the functional at $\kappa_l=0$) and (b) that {\it only} the
non--standard saddle point $\tilde\Lambda$ contributes to the
correlation function. Indeed, for any other non-standard saddle point
one or several signatures corresponding to energy arguments
$\epsilon_n\not= \epsilon_{n_1,n_2}$ are changed. These energies are
not coupled to sources. Repeating the steps outlined above one finds
that the $B$-integral around these saddle points gives zero
(i.e. the $\kappa\to 0$ limit of the factor $\tilde F$ above).  The
same argument also shows that the non-standard saddle points do not
contribute to $Z[0]$.  Differentiating Eq.~(\ref{eq:5}) w.r.t.
$\kappa_{1,2}$ and adding the contribution of the standard saddle
point, we obtain the well known result\\[-0.5cm]
\begin{equation}
\label{eq:6}
R(\omega)=-{1\over2} {\,\rm Re\,}
{1-
\exp(2is) \over s^2}=
  - \left({\sin s\over s}\right)^2 
\end{equation}\\[-0.4cm]
for the two point correlation function of the zero mode theory.
We finally mention that the inclusion of
spatially fluctuating modes into the formalism (a) does not change the 
saddle point structure and (b), after Gaussian integration, leads to a
renormalization $\tilde F\to {\cal
  D}(\omega)\tilde F$, where\\[-0.6cm]
\begin{equation}
\label{eq:7}
{\cal D}(\omega) \equiv \prod\limits_{{\bf q}\neq 0} 
{(D{\bf q}^2)^2 \over (D{\bf q}^2)^2 + \omega^2}\, 
\end{equation}\\[-0.4cm]
and ${\bf q}$ are the quantized momenta associated to fluctuations in
a finite size system.  Combining this with the contribution of
$\Lambda_0$, we reproduce  the familiar result
\cite{Andreev95,Bogomolny96,Kamenev99} for the level statistics of the
unitary disordered electron gas.  Since this result was obtained in
a saddle point approximation, its validity is restricted to energies
$\omega \gg \Delta$. Indeed, in the opposite limit the fluctuation
modes $B$ become too light to be treated in the Gaussian
approximation. The fact that the zero mode result Eq.~(\ref{eq:6}) is
actually exact for any $\omega$ is a ``coincidence'' (see
however Ref.~\cite{Zirnbauer99}), specific to the unitary ensemble.


{\it Part II:} The analysis above crucially relied on the causality
{\it postulate}: summations over continuous functions of $\epsilon$
produce zero.  As mentioned above, this feature could not be 
made explicit  because the validity of the action (\ref{eq:1}) is limited to 
low energies.  In this second part of the
Letter we are going to construct an energetically enlarged formulation
whereby large and small $\epsilon$ are treated on the
same footing. This will make the causality property manifest. At
the same time we will see why the pragmatic scheme employed above is
sufficient for the calculation of low energy observables. 

In order to not unnecessarily complicate the discussion we will
formulate this part of the analysis for an $N$-dimensional random
matrix theory (RMT) Hamiltonian $H=\{H_{\mu\nu}\}$ defined through the
Gaussian correlation law $\langle H_{\mu\nu}\rangle = 0$ and $\langle
H_{\mu\nu} H_{\nu'\mu'}\rangle = N^{-1} \delta_{\mu \mu'} \delta_{\nu
  \nu'}$.  The advantage gained is that $H$ has neatly defined
universal large energy properties, i.e. that we will not  need to
consider the non-universal UV asymptotics of the free--electron
Hamiltonian underlying Eq.~(\ref{eq:1}). Later on we will argue that,
as far as the present discussion is concerned, the specific modeling
of the Hamiltonian is of no relevance. Further, to deal with a
manifestly $UV$-regularized model, we compactify our energy
variables. This can be done by discretizing the temporal Keldysh
contours ${\cal C}^l$ to a lattice of small spacing $\delta_t$.  As
a result, the energy variables $\epsilon_n \in \{
\delta_\epsilon,2\delta_\epsilon,\dots, K\delta_\epsilon\}$, where $K
\equiv 2\pi/(\delta_\epsilon \delta_t) \gg 1$.

The effective Keldysh action for the RMT model can be obtained by a
straightforward adaptation of previous derivations of the 
$\sigma$--model for RMT Hamiltonians 
\cite{HAW} to the specifics of the Keldysh
$\sigma$-model. As an intermediate result one obtains the partition
function $Z[\hat \kappa] =  \int {\cal D}Q
\exp\{iS[Q,\hat \kappa]\}$ where  
the action $iS[Q,\hat\kappa] = -N \left( {1\over 2} {\,\rm tr\,}(Q^2)
  - {\,\rm tr\, ln\,}\left[\hat z + Q\right]\right)$ with $\hat z =
\sigma_3 \delta_t^{-1}( e^{-i\sigma_3 \delta_t \hat \epsilon} -1)$
($\sigma_3$ acts in the Keldysh $2\times 2$ space).  At this stage
$Q=\{Q_{\epsilon,\epsilon'}^{l,l'}\}$ is a $2K$-dimensional matrix
that has been introduced to decouple the $H$-averaged action 
\cite{HAW} (no constraint $Q^2=\openone$ as yet). The unusual phase--type
appearance of the energy argument is due to the time
discretization.  However, in the limit $\epsilon\delta_t \ll 1$, $\hat
z \to -i\hat \epsilon$ and we retrieve the standard form of the RMT
$\sigma$-model action  
\cite{HAW}. We next subject the action to a saddle point
analysis (stabilized by the parameter $N\gg 1$).  Variation of the action
w.r.t. $Q$ yields the quadratic equation $Q = (\hat z + Q)^{-1}$. The
$2^{2K}$--fold degenerate energy diagonal set of solutions, $\Lambda^{(l)}(\epsilon_n)
\equiv \Lambda^{l,l}_{\epsilon_n,\epsilon_n}$, is given by\\[-0.6cm]
\begin{equation}
  \label{eq:8}
\Lambda^{(l)}(\epsilon_n) =-{1\over 2}\Big[ 
z_l(\epsilon_n) \pm i (-z_l^2(\epsilon_n) -4)^{1/2}\Big].  
\end{equation}\\[-0.6cm]
This expression determines the entire spectral structure of the model.
First, notice that for energies $\epsilon_n\ll 1 \ll \delta_t^{-1}$,
$\Lambda^{(l)}(\epsilon_n) = \pm 1+ {\cal O}(\epsilon_n)$. This means
that the solutions $\Lambda^{(l)}(\epsilon_n)$ represent an UV
extension of the saddle points $\Lambda$ discussed in part I. We next
ask whether the two sign alternatives in Eq.~(\ref{eq:8}) are equivalent or
whether the model has a preferred choice. Indeed, the latter is the
case: for energies $\epsilon_n\gg 1$ greatly in excess of the width of
the spectrum, $\Lambda^{(l)}(\epsilon_n)$ must approach zero -- the
free Gaussian saddle point of the non--disordered model. This
condition determines $\Lambda_0(\epsilon_n)= {1\over 2}[-\hat
z(\epsilon_n) + \sigma_3 (-\hat z^2(\epsilon_n) -4)^{1/2}]$ as the
canonical solution. For low energies $\Lambda_0(\epsilon_n)$ reduces to the saddle point
$\Lambda_0$ discussed above.  This saddle point has the 
important  property,
$\sum_n [\Lambda_0(\epsilon_n)]^k=0$, $k$ a positive
integer. The outline of the proof is as follows: (due to the presence of
a finite imaginary increment) the summation over $\epsilon_n$ is
equivalent to an integration of the variable $w=\exp\{-i\delta_t
\epsilon_n\}$ over the complex unit circle. It is straightforward to
verify that for $|w|\ge 1$, $\Lambda_0(w)$ is analytic (the branch cut
singularity of the square root lies {\it inside} the unit circle).
Further, for $|w| \gg 1$, the integrand decays as $w^{-(k+1)}$. From
Cauchy's theorem we conclude that the summation gives zero $\Box$.
Summarizing, we have found a complex saddle point structure which
extends the low energy saddle points $\Lambda =\pm 1$ discussed in
part I into the UV regime. The complex structure of the theory entails
the existence of a 'natural' saddle point $\Lambda_0$.  Existence and
behavior of $\Lambda_0$ vitally depend on the large energy, $\epsilon\gg 1$, 
asymptotics of the theory.

To more explicitly establish contact with the low energy regime
discussed in part I, we next introduce fluctuations around the saddle
points (\ref{eq:8}). Defining $Q= \Lambda + P$, where $P$ is some
Hermitian matrix\cite{foot2}, and expanding the action $S[\Lambda +
P]$ to  second order in $P$ we obtain $iS^{(2)}_\Lambda[P] =
-{N\over 2}{\,\rm tr\,}(P^2 + P \Lambda P \Lambda)$. This is the
UV-extension of the low energy action discussed in part I. Indeed,
substituting the small $\epsilon$ asymptotics of $\Lambda$ and using
that for the RMT model, $\Delta = \pi /N$
\cite{HAW}, we find that for
energies $\epsilon\ll 1$, $S^{(2)}_{\Lambda}$ reduces to the action,
Eq.~(\ref{eq:2}).

One can now step by step repeat the analysis that led to the 
correlation function of part I.  The only difference is that instead of
energy denominators $\sim (\epsilon_n^+ - \epsilon_{n'}^-)$
constructions like $1+\Lambda^{(1)}_0 (\epsilon_n) \Lambda^{(2)}_0
(\epsilon_{n'})$ appear.  Due to the compact phase-type appearance of
the energy arguments in $\Lambda_0(\epsilon_n)$ all energy summations
converge. Further, the properties of $\Lambda_0$ discussed above imply the
vanishing of expressions like $\sum_n f(\Lambda_{0}(\epsilon_n))$
where $f$ may be any analytic function.  This implements the causality
principle.  In parentheses we note that the detailed execution of this
program yields a unit normalization of the partition function, without
{\em any} undetermined prefactors.

The discussion above provokes the obvious question whether the 
conclusions drawn from the high energy asymptotics of the action
are specific to the random matrix model? 
We believe that the answer is negative. Recapitulating the
sequence of arguments, it is evident that everything hinges on the absence
of singularities outside the complex unit circle defined through the
compactified energy argument. This in turn is a guaranteed feature as
long as the single particle retarded SCBA Green function of the model
system has a well defined pole structure below the real
axis. In practice, for condensed matter systems with non-universal
high energy behavior, it may be difficult to find closed solutions of
the mean field equations that manifestly display this feature. We
believe, however, that this is a practical rather than a principle
difficulty. 
Summarizing, the main goal of part II was to show that the causality feature
underlying this and previous analyses of the
effective Keldysh action can be made an explicit ingredient of the
model, on the expense of including large energy asympotics.  
In practical applications of the formalism, one will normally use
causality feature pragmatically, as exemplified in part I. Yet it cannot
be excluded that situations arise, where large scale spectral
structures become essential.

To conclude, on the example of level statistics, we have demonstrated
how non--perturbative quantum effects may be incorporated into the
framework of the dynamic Keldysh $\sigma$--model. In many respects,
the Keldish scheme appears to be simpler and more transparent
than its relatives, replica and SUSY. We expect our methods
to be useful in the analysis of interaction phenomena in disordered
electronic systems.

We benefited a lot from numerous discussions with A. Andreev and M.
Janssen.  A.K. was partially supported by the BSF-9800338 grant.

\end{multicols}

\vspace{-0.5cm}




\begin{thebibliography}{99}

\vspace{-1.6cm}

\bibitem{Efetov}K. B. Efetov, Adv. Phys. {\bf 32}, 53 (1983); 
K. B. Efetov, {\em Supersymmetry in Disorder and Chaos}, 
Cambridge University Press, 1997.  



\bibitem{Wegner79} F.~J.~Wegner, Z. Phys. B {\bf 35}, 207 (1979); K.
  B. Efetov, A.~I.~Larkin, and D.E. Khmelnitskii, Sov. Phys. JETP {\bf
    52}, 568 (1980); A.~M. Finkel'stein, Sov. Phys. JETP {\bf 57}, 97
  (1983); 
  D. Belitz  and T.~R.~Kirkpatrick, Rev. Mod. Phys. {\bf 66}, 261
  (1994).


\bibitem{Kamenev98}
M.~L.~Horbach and G.~Sch\"{o}n, Ann. Phys. {\bf 2}, 51 (1993);
A. Kamenev and A. Andreev, Phys. Rev. {\bf B 60},
  2218, (1999); C.~Chamon, A.~W. W. Ludwig, and C.~Nayak, Phys. Rev.
  {\bf B 60}, 2239, (1999). 
  

\bibitem{Andreev95}A.~V.~Andreev and B.~L.~Altshuler, 
Phys. Rev. Lett.  {\bf 75},  902  (1995); A.~V.~Andreev, B.~D.~Simons, 
and B.~L.~Altshuler, J. Math. Phys. {\bf 37},  4968  (1996).

\bibitem{Kamenev99}A. Kamenev and M.~M\'ezard, J. Phys. A, {\bf 32},
  4373 (1999); Phys. Rev. {\bf B 60}, 3944, (1999); I.~V.~Yurkevich,
  and I.~V.~Lerner, Phys. Rev. {\bf B 60}, 3955, (1999).

\bibitem{Kamenev00}A. Kamenev, cond-mat/0002385. 

\bibitem{ak} A. Altland and A. Kamenev, to be published. 

\bibitem{Bogomolny96}E.~B.~Bogomolny and J.~P.~Keating, 
Phys. Rev. Lett.  {\bf 77},  1472  (1996).

\bibitem{foot1} Within non--interacting problems the Keldysh saddle
  point block $\Lambda^{12}(\epsilon)$ (the distribution function) can
  be chosen at will. Without lost of generality we set
  $\Lambda^{12}(\epsilon)=0$.

\bibitem{Zirnbauer99} M.~R.~Zirnbauer, cond-mat/9903338. 

\bibitem{foot1a} Notice that contractions between $B_{n_1n}$
and $B^\dagger_{n n_1}$ do not contribute. These terms lead to 
denominators $\propto (\epsilon_{n_1}^+ - \epsilon_{n}^-)^{-1}$ which
vanish after summation over $n$ (causality).

\bibitem{HAW} T. Guhr et al., Phys. Rept. {\bf 299}, 189 (1998).

\bibitem{foot2} We here write $\Lambda + P$ (instead of the
  common rotation ansatz $T\Lambda T^{-1}$)
  because to correctly describe the large energy structure of the
  model, fluctuation modes changing the eigen{\it values} of the
  saddle points ('massive modes' in the terminology of the
  $\sigma$--model) must be included.






\end{thebibliography}
\end{document}